\documentclass[preprint,aps,tightenlines,showkeys,nofootinbib,superscriptaddress]{revtex4-2}
\usepackage{color}
\usepackage{enumerate}
\usepackage{latexsym}
\usepackage{graphicx}
\def\be{\begin{equation}}
\def\ee{\end{equation}}
\def\bea{\begin{eqnarray}}
\def\eea{\end{eqnarray}}

\begin{document}

\today\\

\title{Deflection of light by a Coulomb charge in Born-Infeld electrodynamics }

\author{Jin Young Kim \footnote{E-mail address: jykim@kunsan.ac.kr} }
\affiliation{Department of Physics, Kunsan National University,
Kunsan 54150, Korea}

\begin{abstract}
We study the propagation of light under a strong electric field in Born-Infeld electrrdynamics. 
The nonlinear effect can be described by the effective indices of refraction. 
Because the effective indices of refraction depend on the background electric field, the path of light can be bent when the background field is non-uniform. We compute the bending angle of light by a Born-Infeld-type Coulomb charge in the weak lensing limit using the trajectory equation based on geometric optics. We also compute the deflection angle of light by the Einstein-Born-Infeld black hole using the geodesic equation and confirm that the contribution of the electric charge to the total bending angle agree. 

\end{abstract}



\maketitle

\newpage

\section{Introduction}

In Maxwell's electrodynamics the equations of motion for the electromagnetic field are linear and the speed of light is constant in vacuum. The path of light is not affected by the presence of electric or magnetic field. 
However, in quantum electrodynamics, nonlinear trems appear in the effective action from the vacuum polarization effect. Then the path of light can be bent when the light passes around a strong electric or magnetic field. In this sense, vaccum itself can be considered as a dielectric medium in nonlinear electrodynamics. The one-loop correction is described by the Euler-Heisenberg Lagrangian \cite{HeisenbergEuler,Schwinger}. In spite of several ground laboratory experiments to observe the nonlinearity in the presence of a strong magnetic field, the proof is still lackng \cite{BRST,PVLAS,BMV}. 

When photons pass close to compact astronomical objects like a charged black hole or a magnetized neutron star, the path of light is affected by the electromagnetic effect as well as by the gravitational effect. The gravitational lensing by the Reissner-Nordstrom (RN) black hole is a typical example \cite{Weinberg,Chandrasekhar}. Recently, inspired by the development in string theory, interests in the Born-Infeld (BI) theory were renewed. The black hole solution in the Einstein-Born-Infeld (EBI) gravity, which is the nonlinear electromagnetic generalization of RN black hole, is one of the interesting topics. Diverse aspects of the propagation of light in Born-Infeld-type nonlinear electrodynamics were studied  \cite{Plebanski,GibbRash,Aiello,Kruglov07,Kruglov10,Kruglov17}. 

Most ground laboratory experiments and astronomical observations related to the nonlinearity of electromagnetism utilize the magnetic field. It seems natural to think utilizing the electric field to test such nonlinearity. One can consider the case when photons pass very close to heavy nuclei in ground laboratory experiment. In this case it is known that the relevant electric field is greater than $ 10^{21} {\rm V/m }$ \cite{Jackson}. One can also consider the gravitational lensing of light by charged black holes. 
It seems not probable to observe charged black holes because the observed universe is charge neutral. Nonetheless we think it is interesting to think the propagation of light under the strong electric field by a Coulomb charge in Born-Infeld electrodynamics. It may also deepen our understanding when we study the nonlinearity using magnetic field. 

In Born-Infeld-type nonlinear electrodynamics, the propagation of light in a uniform external electromagnetic field can be described by the effective index of refraction \cite{Kruglov07,Kruglov10,Kruglov17}. When the external field is non-uniform, for example the electric field by a spherically symmetric charge distribution, the effective index of refraction can change continuously. Then the path of light incoming from infinity can be bent by the gradient of the effective index of refraction generated by the background field. In this paper we compute the deflection angle of light by a Born-Infeld-type Coulomb charge. 

The organization of this paper is as follows. In Sec. II, we study the propagation of light under the background electric field in Born-Infeld electrodynamics. We calculate the effective indices of refraction for the uniform background electric field. In Sec. III, we compute the bending angle of a light ray passing the strong electric field by a Born-Infeld-type Coulomb charge using the trajectory equation based on geometric optics. In Sec. IV, we compute the bending angle of the Einstein-Born-Infeld black hole in the weak field limit using the geodesic equation. We confirm that the two results agree in the appropriate limit. Finally in Sec. V, we discuss our results.  

\section{Index of refraction in the background electric field} 

The generalized Born-Infeld action, which combines the classical Born-Infeld action \cite{Born, BornInfeld} and 
the one-loop correction of quantum electrodynamics \cite{HeisenbergEuler}, can be described by 
the effective Lagrangian density  \cite{Kruglov10}
\be
 {\cal L}_{eff} = \beta^2 \left ( 1- \sqrt{ 1 + \frac{2 S}{\beta^2} - \frac{P^2}{\beta^2 \gamma^2} } \right ).
 \label{lagrangianeffective}
\ee
Here $S$ and $P$ are two Lorentz-invariants given by 
\be
 S = \frac{1}{4} F_{\mu \nu} F^{\mu \nu} = \frac{1}{2} ( {\bf B}^2 -  {\bf E}^2 )   , 
 ~~~ P = \frac{1}{4} F_{\mu \nu} {\tilde F}^{\mu \nu} =  {\bf E} \cdot {\bf B} ,
 \label{defSP}
\ee
where $F_{\mu \nu} = \partial_\mu A_\nu - \partial_\nu A_\mu $ is the field strength tensor 
and ${\tilde F}_{\mu \nu}  = \frac{1}{2} \epsilon_{\mu \nu \alpha \beta} F^{\alpha \beta} $ is its dual tensor. 
The parameter $\beta$ and $\gamma$ are given by 
\be \frac{1}{ \beta^2} = \frac{1}{ B^2} + 8 a_{\rm QED}, ~~~
       \frac{1}{ \gamma^2} = \frac{1}{ B^2} + 2 b_{\rm QED}, 
 \label{betagamma}
\ee      
where $B$ is the parameter characterizing the maximum value of  field strength in classical Born-Infeld electrodynamics, $a_{\rm QED}$ and $b_{\rm QED}$ are the quantum electrodynamic corrections defined by, in terms the fine structure constant $\alpha$ and the electron mass  $m_e$,
\be a_{\rm QED} = \frac{2}{45}\frac{\alpha^2}{m_e^4}, ~~~ b_{\rm QED} =\frac{14}{45}\frac{\alpha^2}{m_e^4}.
 \label{aandb}
\ee      
When $S / \beta^2$ and $P^2 / \beta^2 \gamma^2$ are small, Eq.  (\ref{lagrangianeffective}) can be expanded as 
\be
 {\cal L}_{eff} \simeq - S  + \frac{S^2}{2\beta^2} +  \frac{P^2}{2\gamma^2} .
 \label{leffapprox}
\ee
In the limit $ B \to \infty$ this gives the Euler-Heisenberg Lagrangian. 

From the Euler-Lagrange equation for vector field $A_\mu$, we have 
\be 
\partial_\mu \left [ \frac{1} {\cal R}  \left ( F_{\mu \nu} - \frac{P}{\gamma^2} {\tilde F}_{\mu\nu} \right ) \right ] = 0 ,
\label{eulerlageq}
\ee
where 
\be
 {\cal R} = \sqrt{ 1 + \frac{2 S}{\beta^2} - \frac{P^2}{\beta^2 \gamma^2} } .
 \label{defcalR}
\ee
Also from the Bianchi identity for field strength tensors, we have
\be 
\partial_\mu  {\tilde F}^{\mu\nu}  = 0 .
\label{bianchiid}
\ee
Eqs. (\ref{eulerlageq}) and (\ref{bianchiid}) can be considered as the generalized Maxwell equations. 
The electric displacement and magnetic field are given from Eq. (\ref{lagrangianeffective})
\bea
{\bf D } &=& \frac {\partial {\cal L}_{eff} }{ \partial {\bf E} }
              = \frac {1 }{\cal R} \left (  {\bf E} + \frac {1 }{\gamma^2} {\bf B} ( {\bf E} \cdot {\bf B} ) \right ),   \label{defofD} \\
{\bf H } &=&- \frac {\partial {\cal L}_{eff} }{ \partial {\bf B} }
              = \frac {1 }{\cal R} \left (  {\bf B} - \frac {1 }{\gamma^2} {\bf E} ( {\bf E} \cdot {\bf B} )\right) ,  \label{defofH}
\eea   
where $ {\bf E}_i = F_{0i}$ and $ {\bf B}_i = - (1/2) \epsilon_{ijk} F^{jk}$. 
From Eqs. (\ref{eulerlageq}) and (\ref{bianchiid}),  Maxwell equations can be written  in terms of $ {\bf D}, {\bf B}, {\bf E}$,  and ${\bf H}$ as
\bea
\nabla \cdot {\bf D} &=& 0 , ~~~~\frac{\partial {\bf D} } {\partial t } - \nabla \times {\bf H } = 0 ,   \nonumber   \\
\nabla \cdot {\bf B} &=& 0 , ~~~~\frac{\partial {\bf B} } {\partial t } + \nabla \times {\bf E } = 0  .
\label{genMaxwelleqs}
\eea

From the definition $ D_i = \varepsilon_{ij} E_j$ and $ B_i = \mu_{ij} H_j$, the electric permittivity tensor and the inverse magnetic  permeability tensor are obtained as 
\be
\varepsilon_{ij} = \frac {1 }{\cal R} \left (  \delta_{ij} + \frac {1 }{\gamma^2} B_i B_j  \right ),   ~~~
(\mu^{-1} )_{ij}  = \frac {1 }{\cal R} \left (  \delta_{ij} - \frac {1 }{\gamma^2} E_i E_j  \right ). \label{emutensors}
\ee
The above equations show that the vacuum is anisotropic in the generalized Born-Infeld electrodynamics. 
The eigenvalues and the inverse matrix which diagonalizes $\varepsilon_{ij}$ are obtained as 
\bea 
\lambda_1  (\varepsilon ) &=& \frac{1}{\cal R} \left ( 1 + \frac{ {\bf B}^2 }{\gamma^2} \right ), ~~
\lambda_2  (\varepsilon ) = \frac{1}{\cal R} ~({\rm degenerate}),   \\
(\varepsilon^{-1} )_{ij} &=& {\cal R} \left ( \delta_{ij} - \frac{B_i B_j}{ \gamma^2 + {\bf B}^2 } \right ) , 
\eea
The corresponding values for $\mu_{ij}^{-1}$ are 
\bea 
\lambda_1  (\mu^{-1} ) &=& \frac{1}{\cal R} \left ( 1 - \frac{ {\bf E}^2 }{\gamma^2} \right ), ~~
\lambda_2  (\mu^{-1}  ) = \frac{1}{\cal R} ~({\rm degenerate}),   \\
\mu_{ij} &=& {\cal R} \left ( \delta_{ij} + \frac{E_i E_j}{ \gamma^2 - {\bf E}^2 } \right ) . 
\eea

Now we consider the propagation of the plane electromagnetic wave  $({\bf e} ,{\bf b})$ under the constant uniform background electric field
\be
{\bf e} = {\bf e}_0 e^{i ( {\bf k} \cdot  {\bf r} - \omega t) }, ~~{\bf b} = {\bf b}_0 e^{i ( {\bf k} \cdot  {\bf r} - \omega t) }. 
\ee
We take the direction of the background electric field as $x$-direction ${\bar {\bf E}}= ( {\bar E} , 0, 0 ) $ and the direction of propagating wave as $z$-direction. We consider the case where the background electric field is much stronger than the photon's electric and magnetic fields. 
The electromagnetic fields are the superposition of the traveling wave and the background field. 
Substituting ${\bf E} = {\bf e} + {\bar {\bf E} } $ and  ${\bf B} = {\bf b}$ in Eq. (\ref{lagrangianeffective}), up to the quadratic order in ${\bf e}$ and ${\bf b}$,
the effective Lagrangian is given by
\be
 {\cal L}_{eff} ( {\bf e} + {\bar {\bf E} },  {\bf b} )
 = \beta^2 \left ( 1- \sqrt{ 1 + \frac{ {\bf b}^2 - ( {\bf e} +  {\bar {\bf E } })^2 }{\beta^2} 
 - \frac{ ( {\bf b} \cdot { \bar {\bf E }} )^2} {\beta^2 \gamma^2} } \right ).
 \label{leffsum}
\ee
The electric displacement and the magnetic field are obtained as
\bea
d_i &=& \frac {\partial {\cal L}}{\partial e_i} = \frac {1}{\kappa} \left ( \delta_{ij} 
+ \frac{{ \bar E}_i {\bar E}_j }{ \beta^2 \kappa^2} \right ) e_j ,  \\
h_i &=& - \frac {\partial {\cal L}}{\partial b_i} = \frac {1}{\kappa} \left ( \delta_{ij} 
- \frac{{ \bar E}_i {\bar E}_j }{ \gamma^2 } \right ) b_j , 
\eea
where 
\be 
\kappa =  \sqrt{ 1 - \frac{ {\bar {\bf E}}^2 }{\beta^2} }. 
\label{kappa}
\ee
The electric permittivity and the inverse magnetic  permeability tensors are read as
\bea
\varepsilon_{ij} &=& \frac {1}{\kappa} \left ( \delta_{ij} 
+ \frac{{ \bar E}_i {\bar E}_j }{ \beta^2 \kappa^2} \right )  , ~~~\label{permittivitytensor} \\
(\mu^{-1})_{ij} &=& \frac {1}{\kappa} \left ( \delta_{ij} 
- \frac{{ \bar E}_i {\bar E}_j }{ \gamma^2 } \right ) .     \label{permeabilitytensor}
\eea
The above tensors for the constant uniform electric field can be obtained from those for constant uniform magnetic field \cite{Kruglov10} by replacing $ {\bar {\bf B} } $, $\gamma^2$, $\beta^2 \kappa^2$, 
$\kappa =  \sqrt{ 1 + { {\bar {\bf B}}^2 }/{\beta^2} }$ 
with
$ {\bar {\bf E} } $, $\beta^2 \kappa^2$, $\gamma^2$, 
$\kappa =  \sqrt{ 1 - { {\bar {\bf E}}^2 }/{\beta^2} }$, respectively. 

The Maxwell equations in (\ref{genMaxwelleqs}) can be written in terms of ${\vec k} $ and $\omega$ as
\be 
k_i d_i = k_i b_i = 0 , ~~~ {\bf k} \times {\bf e} = \omega  {\bf b} , ~~~{\bf k} \times {\bf h} = - \omega  {\bf d} , 
\label{maxeqskandomega}
\ee
and the wave equation for the electric field ${\bf e}$ is obtained as 
\be 
\left [ {\bf k}^2 ( \mu^{-1} )_{ji} + \{ k_a   ( \mu^{-1} )_{al} k_l - {\bf k}^2 ( \mu^{-1} )_{aa} \} \delta_{ij} 
-  k_l ( \mu^{-1} )_{jl}  k_i + \omega^2 \varepsilon_{ij} \right ] e_j = 0.  
\label{waveeqn} 
\ee
Substituting Eqs. (\ref{permittivitytensor}) and (\ref{permeabilitytensor}) in Eq. (\ref{waveeqn} ) and using 
${\bf k} \cdot {\bf e} = {\bf k} \cdot {\bar {\bf E}} = 0$, we obtain
\be 
\frac{\omega^2}{\kappa} \left [ \left ( 1 - n^2 + n^2 \frac{ {\bar {\bf E}}^2 }{\gamma^2} \right)  \delta_{ij} 
+ \left(  \frac{1}{ \beta^2 \kappa^2} - \frac{n^2} {\gamma^2}  \right ) {\bar E}_i {\bar E}_j \right ] e_j = 0 ,
\label{waveeqn2} 
\ee
where $n= k / \omega$ is the index of fraction.  The condition for Eq. (\ref{waveeqn2}) to have nontrivial solutions is that 
the determinant of the following matrix is zero
\be 
\Lambda_{ij} =  \left ( 1 - n^2 + n^2 \frac{ {\bar {\bf E}}^2 }{\gamma^2} \right)  \delta_{ij} 
+ \left(  \frac{1}{ \beta^2 \kappa^2} - \frac{n^2} {\gamma^2}  \right ) {\bar E}_i {\bar E}_j . 
\label{Lambdamatrix} 
\ee
The eigenvalues of the matrix $\Lambda_{ij}$ are obtained as
\be 
\lambda_1 =  1 - n^2 + n^2 \frac{ {\bar {\bf E}}^2 }{\gamma^2} , ~~~
\lambda_2 =  1 - n^2 + \frac{ {\bar {\bf E}}^2 }{\gamma^2 \kappa^2} .
\label{eigenvalofLambda} 
\ee
Using $\kappa =  \sqrt{ 1 - {\bar {\bf E}}^2 /{\beta^2} }$, the effective indices of refraction are obtained, from $\lambda_1 =0$ and  $\lambda_2 =0$, as
\be 
n_{\bot} =  \left ( 1 - \frac{ {\bar {\bf E}}^2 }{\gamma^2} \right)^{-\frac{1}{2}}  , ~~~
n_{\parallel} =  \left ( 1 - \frac{ {\bar {\bf E}}^2 }{\beta^2} \right)^{-\frac{1}{2}}. 
\label{indexofrefraction} 
\ee
Note that there is the vacuum birefringence effect for $\beta \ne \gamma$, coming from the QED corrections. 

\section{Deflection angle by geometric optics}
When light passes a uniform constant background electric field, the propagation of light can be described by effective indices of refraction which depend on the background electric field. 
A non-uniform background electric field can make gradients for indices of refraction. 
If the light ray incoming from infinity passes a non-uniform electric field, the path of light can be bent by the gradients. The bending angle can be computed by calculus. 

The simplest non-uniform and isotropic gradient can be made by a Coulomb charge or a spherically symmetric charge distribution. We consider the BI electrostatic case with ${\bf B} = {\bf H}= 0$ assuming a point charge $Q$ is located at the origin
\be 
\nabla \cdot {\bf D} = Q \delta ({\bf r}) .
 \ee 
The electric displacement is obtained as 
\be 
 {\bf D} = \frac{ Q }{4 \pi r^2} {\hat r} ,
 \label{displacementBI}
 \ee
and, from Eq. (\ref{defofD}), the electric field is obtained as
\be 
 {\bf E} = \frac{ Q }{4 \pi \sqrt{r^4 + \rho_0^4 } } {\hat r} ,    
 \label{electricfield}
 \ee
where $\rho_0 = \sqrt{ |Q| / 4 \pi \beta}$. 

If we denote the unit vector in the direction of light by ${\bf u}={\bf k}/k$ as in \cite{Bialynicka,Adler,HeylHern,GitGies}, the indices of refraction can be written as 
\be 
n_{\bot} =  \left ( 1 - \frac{ ( {\bar {\bf E} } \times {\bf u} )^2 }{\gamma^2} \right)^{-\frac{1}{2}}  , ~~~
n_{\parallel} =  \left ( 1 - \frac{  ( {\bar {\bf E} } \times {\bf u} )^2   }{\beta^2} \right)^{-\frac{1}{2}}. 
\label{nCoulombic} 
\ee
The maximum field strength is $E_{\max} =\beta$ from Eq. (\ref{electricfield}). As mentioned before the lower bound of the classical BI parameter $B$ in Eq. (\ref{betagamma}) is of the order $B > 10^{21} {\rm V/m}$  while $a_{\rm QED}$ and $b_{\rm QED}$ are of the order $E_c^{-2}$, where $E_c$ is the QED critical field stregth defined by $E_c = {m_e^2 c^3}/e\hbar =1.32 \times 10^{18} {\rm V/m}$. In the limit $B \to \infty$, the bending of light comes purely from QED one-loop effect described by Euler-Heisenberg Lagrangian \cite{KimLee1,KimLee1}. 
When the light ray is passing the region where the field strength $E$ is comparable to the BI parameter $B$, one can compute the trajectory numerically.
We consider the case when the light ray is passing the region where $E$ is not as large as $B$. This corresponds to the weak field limit. The leading terms in Eq. (\ref{nCoulombic}) are 
\be 
n_{\bot} =   1 +  \frac{ ( {\bar {\bf E} } \times {\bf u} )^2 }{2 \gamma^2}   , ~~~
n_{\parallel} =  1 + \frac{ ( {\bar {\bf E} } \times {\bf u} )^2 }{2 \beta^2}. 
\label{nCoulapprox} 
\ee

Because the electric field by the point charge $Q$ is isotropic, we may assume that the light ray is confined to the equitorial plane $\theta = \pi /2$ and the trajectory of light $y(x)$ is on the $xy$ plane.
If we take the direction of incoming ray at $x = - \infty$ as the $+x$ direction, the slope $y' = dy/dx$ is the direction of the light ray (Fig. 1). Substituting $r = \sqrt{x^2 + y^2 }$,
${\bf u } = (dx, dy)/\sqrt{dx^2 + dy^2} = (1, y^\prime)/\sqrt{1 + y^{\prime 2}} $ and Eq. (\ref{electricfield}) into Eq. (\ref{nCoulapprox}), the indices of refraction can be written as 
 \bea
n_{\bot} &=& 1 + \frac{1}{2} \frac{\beta^2}{\gamma^2} \frac{\rho_0^4}{r^4 + \rho_0^4}
                     \frac{(y - xy^\prime)^2}{ r^2 (1 +y^{\prime 2})} , \label{nperpendicular} \\
n_{\parallel}  &=&  1 + \frac{1}{2}  \frac{\rho_0^4}{r^4 + \rho_0^4}
                     \frac{(y - xy^\prime)^2}{ r^2 (1 +y^{\prime 2})} .
\label{nparallel} 
\eea
\begin{figure}
\begin{center}
\includegraphics[height=10cm,keepaspectratio]{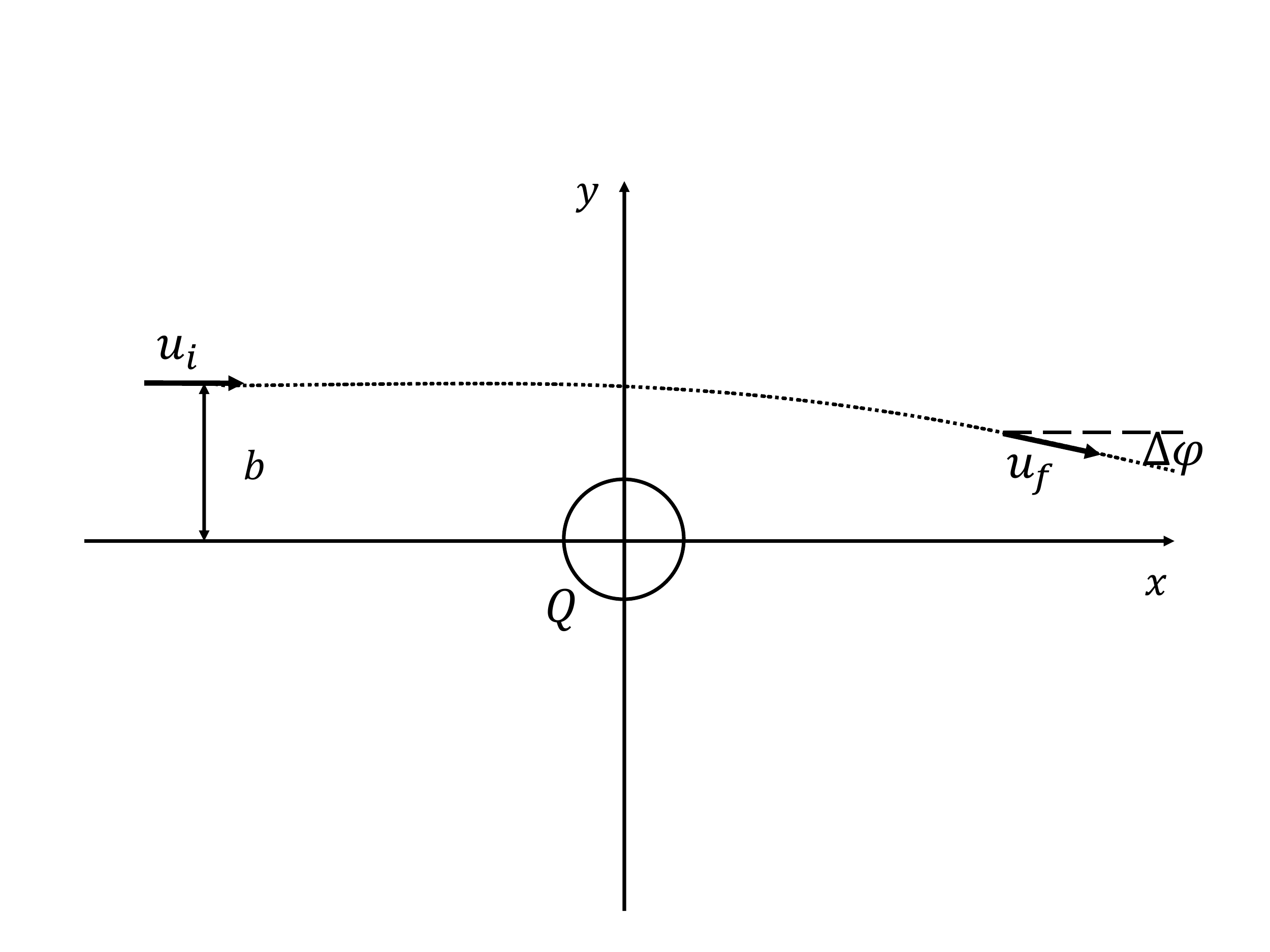}
\caption{Schematic of light bending in a Coulombic field. } \label{fig1}
\end{center}
\end{figure}

When the index of fraction is continually varying, the trajectory of light can be obtained from the following equation derived from Snell's law \cite{KimLee1,KimLee2}
 \be
 \frac{d{\bf u}}{ds}=\frac{1}{n}({\bf u}\times {\nabla}n)\times {\bf u},
 \ee
where $s$ is the distance parameter $ds=|d\vec{\bf r}|$ and ${\bf u}= {d {\bf r}}/{ds}$. 
This trajectory eqaution is equivalent to the eikonal equation of wave optics. For the light bending by the background magnetic dipole field, it has been confirmed that the result using this trajectory equation exactly agree with the result using the eikonal equation \cite{KimLee2, dds}. 
When the index of refraction is close to one, the trajectory equation in the leading order can be approximated as
 \be
 \frac{d{\bf u}}{ds}=( {\bf u}_0 \times \nabla n) \times {\bf u}_0,  \label{trajeceqleading}
 \ee
 where ${\bf u}_0$ denotes the initial direction of the incoming ray.
For the light ray coming from $x=-\infty$ and moving to $+x$ direction, 
 \be
 {\bf u}_0=(1,0,0),
 \ee
and the leading-order trajectory equation (\ref{trajeceqleading}) can be written as
 \be
 \frac{d^2x}{ds^2} = 0, ~~~ \frac{d^2y}{ds^2} = \frac{\partial n}{\partial y} .      \label{trajeq}  
 \ee
From the first equation $ds=dx$ at the leading order and $d^2y / ds^2$ in the second equation can be replaced by  $d^2y / dx^2$. 

Now we compute the deflection angle with the impact parameter $b$. The initial conditions can be written as
 \be
 y(-\infty)=b, ~~~~ y'(-\infty)=0 . \label{inicond}
 \ee
Also the $y^\prime $ in Eqs. (\ref{nperpendicular}) and (\ref{nparallel}) can be neglected in the leading order approximation.  
Then the trajectory equation to compute the deflection angle is
 \be
 y''  = \xi \rho_0^4 \left [ \frac{y}{r^2 (r^4 + \rho_0^4 ) } - \frac{y^3}{r^4 (r^4 + \rho_0^4 ) }- \frac{2y^3}{ (r^4 + \rho_0^4 )^2 } \right] ,
 \label{ytwoprime}
 \ee
 where $\xi =\beta^2 / \gamma^2$ for the perpendicular mode $n_{\bot}$ and $\xi =1$ for the parallel 
mode $ n_{\parallel}$. 
By integrating Eq. (\ref{ytwoprime})
  \be
  y' (\infty) - y'(-\infty) 
 = \xi \rho_0^4 \int_{-\infty}^{\infty} \left [ \frac{y}{r^2 (r^4 + \rho_0^4 ) } - \frac{y^3}{r^4 (r^4 + \rho_0^4 ) } 
 - \frac{2y^3}{ (r^4 + \rho_0^4 )^2 } \right] dx ,
 \label{totalintegral}
 \ee
we can get the deflection angle by 
 \be 
 \Delta \varphi =  y' (\infty) - y'(-\infty) . 
 \ee
 The details of the integration are given in the appendix. Substituting the result in Eq. (\ref{intIfinal}), we obtain the deflection angle as 
 \be
 \Delta \varphi
 = \xi \left[ \frac{\pi}{2} - \frac{\pi}{\sqrt{2}} \frac {b}{ \left( \sqrt{b^4 + \rho_0^4} + b^2  \right)^{1 / 2} }
     - \frac{\pi}{2 \sqrt{2}} \frac {b^3 \rho_0^4 \left( 3 \sqrt{b^4 +\rho_0^4} + 2 b^2 \right )}
                                                       {\left( b^4 +\rho_0^4 \right)^{3/2} \left ( \sqrt{b^4 + \rho_0^4} + b^2 \right )^{3 / 2} }  \right ].
 \label{yprimedifference}
 \ee
 
Let us consider the limiting case $B \rightarrow \infty$ to compare with the known result in the literature. 
Because $\rho_0 (= \sqrt{ |Q|/ 4 \pi \beta} )$ is small when $\beta$ is large, the leading term in Eq.  (\ref{yprimedifference}) is
 \be
 \Delta \varphi
 = - \xi \frac{9 \pi}{16} \frac{\rho_0^4}{b^4} = - \xi \frac{9}{256 \pi} \frac{Q^2}{\beta^2 b^4}.
 \label{yprimediffapp}
 \ee
The negative sign in Eq. (\ref{yprimediffapp}) means that the bending is attractive toward the charge. 
From the definition of parameter $\beta^2 $ in Eqs. (\ref{betagamma}) and (\ref{aandb}), in the limit $B \to \infty$,
\be 
\frac{1}{ \beta^2} \to  \frac{16}{45}\frac{\alpha^2}{m_e^4},
\ee
and Eq. (\ref{yprimediffapp}) reduces to
\be
 \Delta \varphi
 = - \xi \frac{1}{80 \pi} \frac{\alpha^2}{m^4} \frac{Q^2}{ b^4}.
 \label{bendingangleEH}
 \ee
For parallel mode $\xi =1$ and for perpendicular mode mode $\xi = (7/4)^2$.  This result exactly agrees with the calculation from the Euler-Heisenberg Lagrangian \cite{KimLee1}. 
 \section{Deflection angle by geodesic equation}
In the previous section we calculated the bending angle by a Born-Infeld-type Coulomb charge from the trajectory equation based on geometric optics. 
Here we consider the general relativistic effect and compute the bending angle using the geodesic equation. 
We start from the black hole solution in Einstein-Born-Infeld gravity. 
The EBI action is given by
\be
 S = \int d^{4} x \sqrt{-g} \left ( \frac{R }{16 \pi G}  + L_{BI}     \right ),
 \label{ebiaction}
\ee
where the BI Lagrangian density is given by Eq. (\ref{lagrangianeffective}) $L_{BI} = {\cal L}_{eff}$.
For the static electrically-charged case, the solutions of the spherically symmetric black hole are given by \cite{GibbRash,leekimpark} 
 \be
 ds^2 = - f(r) dt^2 + \frac{1}{f(r)} dr^2 + r^2 (d \theta^2 + \sin^2 \theta d \phi^2),
 \label{metricsol}
 \ee
 \be
 {\bf D} = \frac{Q_E }{ r^2} {\hat r}, 
 \label{dinEBI}
 \ee
 where 
 \be 
 f (r) = 1 - \frac{2M}{r} + \frac{2 \beta^2} {r} \int_r^\infty \left ( \sqrt{x^4 + \frac{ Q_E^2}{\beta^2} } - x^2 \right ) dx ,
 \label{psimetric} 
 \ee
$M$ is the ADM mass, and $Q_E$ is the electric charge. The asymptotic form of the metric function in the large-distance limit $r \gg \sqrt{|Q_E | /\beta}$ can be written as
 \be 
 f (r) = 1 - \frac{2M}{r} + \frac{Q_E^2}{r^2} -  \frac{ Q_E^4}{ 20 \beta^2 r^6} +\cdots .
 \label{psimetricasym} 
 \ee
 It is obivous that  the Reissner-Nordstrom metric is recovered in the limit $\beta \to \infty$. 

 Photons do not follow the geodeic of the gravitational field due to the coupling of nonlinear electrodynamics to gravity. They follow the effective null geodesic generated by the self-interaction 
of the nonlinear theory. The effective metric of light is given by \cite{Delorenci, Breton}
  \be
 ds_{eff}^2 = - \chi(r) f(r) dt^2 + \frac{\chi(r)}{f(r)} dr^2 +  r^2 (d \theta^2 + \sin^2 \theta d \phi^2),
 \label{photoneffmetric}
 \ee
where
\be
\chi(r) =\left ( 1 + \frac{Q_E^2}{\beta^2 r^4} \right )^{-1} .
\label{omega}
\ee
We will calculate the deflection angle using this effective metric. 

We consider the motion of a photon in the static spherically symmetric metric of the form, following the notation and procedure in \cite{Weinberg},
 \be
 ds^2 = - B(r) dt^2 + A(r) dr^2 +C(r) (d \theta^2 + \sin^2 \theta d \phi^2). 
 \label{sphsymmetric}
 \ee
The bending angle of light coming from infinity is given by 
\be 
\Delta \varphi = 2 | \varphi (r_0 ) - \varphi_\infty | - \pi, 
 \label{weinberg855}
\ee
whete $r_0$ is the distance of closest approach and 
\be
 \varphi (r_0 ) =  \varphi_\infty + \int_{r_0 }^\infty
    \left [ \frac {A(r)} {C(r)} \right ]^{1/2}    \left [ \frac {C(r) B(r_0 )} {C(r_0 ) B(r)} \right ]^{-1/2} dr .
  \label{weinberg866}
 \ee                        
Substituting Eq.(\ref{photoneffmetric}) in Eq.(\ref{weinberg866}), we have 
\be
 \varphi (r_0 ) - \varphi_\infty = I (r_0 ) = 
 \int_{r_0 }^\infty
    \left [ \frac {\chi(r)} {f(r)} \right ]^{1/2}  \left [ \frac {r^2 \chi(r_0) f(r_0) } {r_0^2  \chi(r) f(r)} \right ]^{-1/2} 
\frac{dr}{r} .
  \label{angleintegral}
 \ee  

In general the bending angle of light by EBI black hole can be calculated from the above integral numerically \cite{Eiroa}. Since we are interested in the bending by purely BI electromagnetic effect to compare with the result in the previous section, we compute the deflection angle of EBI black hole in the weak deflection limit and turn off the mass term $M=0$ at the end. In this limit the $Q_E^4$ term in Eq. (\ref{psimetricasym}) is a subleading term compared with the $Q_E^2$ term. Thus, to the leading order, the metric function $f(r)$ in (\ref{angleintegral}) is the same as the metric function of the RN black hole 
 \be 
 f (r) = 1 -\frac{2M} {r} + \frac{Q_E^2} {r^2} .
 \label{psiapprox} 
 \ee
Substituting Eqs. (\ref{psiapprox}) and (\ref{omega}) in  Eq. (\ref{angleintegral}) and defining 
$x \equiv r_0 /r$, up to the quadratic order in $M$ and $Q_E$, we have
\bea
 I (r_0 ) = \int_{0}^1 \frac{dx}{\sqrt{1-x^2}} 
 \bigg[  1 &+& \frac{M}{r_0}  \left( 1 + \frac{1}{1+x} \right)  
       + \frac{M^2}{r_0^2} \left( \frac{3x}{1+x} + \frac{3}{2 (1+x)^2 } + \frac{3}{2} x^2  \right) 
                                       \nonumber            \\
     &-& \frac{1}{2} \frac{Q_E^2}{r_0^2} (1 + x^2 ) + \frac{1}{2}  \frac{Q_E^2}{\beta^2 r_0^4} (1 + x^2 -x^4 )
   \bigg] .    \label{irzero}  
\eea
The integral yields
 \be
   I (r_0 )  = \frac{\pi} {2} + \frac{2M}{r_0} + \left( \frac{15 \pi}{8}  - 2 \right) \frac{M^2}{r_0^2}
    - \frac{3 \pi}{8} \frac{Q_E^2}{r_0^2} + \frac{9 \pi}{32} \frac{Q_E^2}{\beta^2 r_0^4} .
  \label{irzerofinal}
 \ee  
Inserting Eq. (\ref{irzerofinal}) in Eq. (\ref{weinberg855}), we have
\be 
\Delta \varphi =  \frac{4M}{r_0} + \left( \frac{15 \pi}{4} - 4 \right) \frac{M^2}{r_0^2}
    - \frac{3 \pi}{4} \frac{Q_E^2}{r_0^2} + \frac{9 \pi}{16} \frac{Q_E^2}{\beta^2 r_0^4} .
\label{anglediff}
\ee 

 Computation of higher order terms will be straightforward.
Except for the last term, Eq. (\ref{anglediff}) is the deflection angle for the weak lensing by RN black hole \cite{jusufi}. When the bending is small we can replace $r_0$ with the impact parameter $b$. Also the sign of the last term is the same as the first term which is attractive. This means that the bending is toward the black hole. The third term comes from the general relativistic correction by $f(r) \ne 1$. The calculation in Sec. III  is based on geometric optics formalism which correspond to $f (r) = 1$. So this term does not appear in the previous section. 
Because $Q_E = Q/4 \pi$ comparing Eq. (\ref{dinEBI}) with Eq. (\ref{displacementBI}), the last term in Eq. (\ref{anglediff}) exactly agrees with the result in Eq. (\ref{yprimediffapp}).

 \section{Conclusion}

We have studied the trajectory of a light ray under the background electric field of a Born-Infeld-type Coulomb charge.  The nonlinear effect can be described by the effective indices of refraction. Assuming that the background electric field is much stronger than the photon's electromagnetic field, we computed the bending angle in the weak lensing limit. We calculated the bending angle using the trajectory equation based on geometric optics. In the limit where the classical Born-Infeld parameter $B$ is infinite, our computation yields the bending angle by Euler-Heisenberg action. Then we did the same calculation using the geodesic equation of the Einstein-Born-Infeld black hole. We confirmed that the two results agree in the proper limit. 

In general, the velocity of light depends on the effective nonlinear electromagnetic Lagrangian as well as the effective modification of the flat background by mass and charge. The relative sign of the third term and the fourth term in Eq. (\ref{anglediff}) is different. This means that the contribution to the total deflection angle by nontrivial metric function $f(r) \ne 1$ is repulsive while the contribution by nonlinear electrodynamics is attractive. So the total bending algle of EBI black hole is slightly larger than that of RN black hole. When one uses the magnetic field to test the nonlinearity of electrodynamics, one should also consider such general relativistic correction. It is interesting to study the bending of light by the magnetic dipole field in Born-Infeld electrodynamics. 


\section*{Acknowledgements}
This work was supported by Basic Science Research Program through the National Research Foundation of Korea (NRF) funded by the Ministry of Education, Science and Technology (NRF-2019R1F1A1060409).

\section*{Appendix}
 In this appendix we compute the integral in Eq. (\ref{totalintegral})
\be
  I = \rho_0^4 \int_{-\infty}^{\infty} \left [ \frac{y}{r^2 (r^4 + \rho_0^4 ) } - \frac{y^3}{r^4 (r^4 + \rho_0^4 ) } 
 - \frac{2y^3}{ (r^4 + \rho_0^4 )^2 } \right] dx .
 \label{integalI}
 \ee
In the leading order we can put $y=b$ and $r^2 = x^2 + b^2$ in the integrand
  \be
  I =  \rho_0^4 \int_{-\infty}^{\infty} \left [ \frac{b}{(x^2 +b^2) \{ (x^2 +b^2)^2 + \rho_0^4 \} } 
  - \frac{b^3}{(x^2 +b^2)^2 \{ (x^2 +b^2)^2 + \rho_0^4 \} } 
 - \frac{2b^3}{ \{ (x^2 +b^2)^2 + \rho_0^4 \}^2 } \right] dx .
\ee
By partial fraction of the first two terms, the integral can be written as
  \be
  I = b \int_{-\infty}^{\infty} \left [  \frac{1}{(x^2 +b^2)} -  \frac{b^2}{(x^2 +b^2 )^2} 
   -  \frac{x^2}{(x^2 +b^2 )^2 + \rho_0^4 } -  \frac{2b^2 \rho_0^4 }{\{ (x^2 +b^2 )^2 + \rho_0^4 \}^2 } 
      \right] dx .
      \label{intI}
\ee
The first and second integrals are easily computed as
 \be
  I_1 = b \int_{-\infty}^{\infty}  \frac{1}{x^2 +b^2} dx =  b \left [ \frac{1}{b} \tan^{-1} \left ( \frac{x}{b} \right ) \right ]_{-\infty}^{\infty} = \pi ,
\label{Int1}
\ee
 \be
  I_2 = b^3 \int_{0}^{\infty}  \frac{1}{(x^2 +b^2)^2} dx = b^3 \left [ \frac{x}{2 b^2 (x^2 +b^2 )} 
   + \frac{1}{2 b^3} \tan^{-1} \left ( \frac{x}{b} \right ) \right ]_{-\infty}^{\infty} = \frac{\pi}{2} . 
   \label{Int2}
\ee

We do the third integral using the contour integral
 \be
  I_3 = b \int_{-\infty}^{\infty}  \frac{z^2}{(z^2 +b^2 )^2 + \rho_0^4 }  dz .
\ee
There are four simple poles at 
\be 
z = \pm (b^4 + \rho_0^4 )^{1/4} e^{\pm i \alpha/2}, 
\ee
where 
\be 
\alpha = \tan^{-1} \left ( - \frac{\rho_0^2}{b^2} \right ), ~~~ ( \pi/2  < \alpha < \pi ).
\label{definealpha}
\ee
If we take the contour a semicircle in the upper-half plane, the relevant poles are 
\be 
z_1 =  (b^4 + \rho_0^4 )^{1/4} e^{ i \alpha/2}, ~~~ z_2 = - (b^4 + \rho_0^4 )^{1/4} e^{- i \alpha/2},
\ee
and the corresponding redidues are  
\bea
a_{-1} (z_1 ) &=&  \frac{  e^{ i \alpha/2} } {8 i  (b^4 + \rho_0^4 )^{1/4} \sin ( \alpha/2) \cos ( \alpha/2) } ,  \\
a_{-1} (z_2 ) &=& \frac{  e^{- i \alpha/2} } {8 i  (b^4 + \rho_0^4 )^{1/4} \sin ( \alpha/2) \cos ( \alpha/2) } .
\eea
By the residue theorem, we compute the integral  as
\be
  \int_{-\infty}^{\infty}  \frac{z^2}{(z^2 +b^2 )^2 + \rho_0^4 }  dz  =  2 \pi i [ a_{-1} (z_1 ) + a_{-1} (z_2 ) ]=  \frac{\pi}{2} 
\frac{ 1} { (b^4 + \rho_0^4 )^{1/4} \sin ( \alpha/2) } .   \label{integral3}
\ee
From Eq. (\ref{definealpha}), 
\be
\sin (\alpha /2) =  \frac{1}{\sqrt{2} } \left ( 1 + \frac{b^2} {\sqrt{ b^4 + \rho_0^4} } \right )^{1/2} , 
\ee
and substituting $\sin (\alpha /2)$ into Eq. (\ref{integral3}), we find 
\be
I_3  =  \frac{\pi}{\sqrt{2} }   \frac{b} {\left( \sqrt{ b^4 + \rho_0^4} +b^2 \right)^{1/2} }. 
\label{Int3}
\ee

The last integral can be written as
 \be
  I_4 = 2 b^3 \rho_0^4  \int_{-\infty}^{\infty}  \frac{1}{[ (z^2 +b^2 )^2 + \rho_0^4 ]^2}  dz 
  = 2 b^3 \rho_0^4 \left [ - \frac{\partial}{\partial c}   \int_{-\infty}^{\infty}  \frac{1}{(z^2 +b^2 )^2 + c }  dz \right ]_{c= \rho_0^4}  .
\ee
For the integrand $f(z) = 1/ [ (z^2 +b^2 )^2 + \rho_0^4 ]$, the relevant poles are the same as the poles in ${I_3}$ and the corresponding residues are 
\bea
a_{-1} (z_1 ) &=& \frac{  e^{ -i \alpha/2} } {8 i  (b^4 + \rho_0^4 )^{3/4} \sin ( \alpha/2) \cos ( \alpha/2) } ,  \\
a_{-1} (z_2 ) &=& \frac{  e^{ i \alpha/2} } {8 i  (b^4 + \rho_0^4 )^{3/4} \sin ( \alpha/2) \cos ( \alpha/2) }  ,
\eea
and, from the residue theorem, we have
\be
 \int_{-\infty}^{\infty}  \frac{1}{(z^2 +b^2 )^2 + \rho_0^4 }  dz 
 = \frac{\pi}{\sqrt{2}} \frac{1} { \sqrt{b^4 + \rho_0^4 }  \left( \sqrt{ b^4 + \rho_0^4 } +b^2 \right)^{1/2} } .
 \ee
Using 
\be
-\frac {\partial}{\partial c} \left [ \frac{\pi}{\sqrt{2}} \frac{1} { \sqrt{b^4 + c } \left ( \sqrt{ b^4 + c } +b^2 \right )^{1/2} } \right ]
=  \frac{\pi}{4 \sqrt{2}} \frac {3  \sqrt{b^4 + c } + 2b^2 } { (b^4 + c)^{3/2} \left ( \sqrt{ b^4 + c } +b^2 \right)^{3/2} },
 \ee
we have 
\be
I_4 =  \frac{\pi b^3 \rho_0^4}{2 \sqrt{2}} \frac {3  \sqrt{b^4 + \rho_0^4 } + 2b^2 } 
{ (b^4 + \rho_0^4)^{3/2}  \left( \sqrt{ b^4 + \rho_0^4 } +b^2 \right )^{3/2} }.
\label{Int4} 
\ee
Substituting $I_1 , I_2, I_3$, and $I_4$ in Eq. (\ref{intI}), we finally obtain
 \be
I = \frac{\pi}{2} - \frac{\pi}{\sqrt{2}} \frac {b}{ \left( \sqrt{b^4 + \rho_0^4} + b^2  \right)^{1 / 2} }
     - \frac{\pi}{2 \sqrt{2}} \frac {b^3 \rho_0^4 \left( 3 \sqrt{b^4 +\rho_0^4} + 2 b^2 \right )}
                                                       {\left( b^4 +\rho_0^4 \right)^{3/2} \left ( \sqrt{b^4 + \rho_0^4} + b^2 \right )^{3 / 2} }  .
 \label{intIfinal}
 \ee

\end{document}